\documentclass[twoside]{article}
\usepackage{fleqn,espcrc2} 
\usepackage{epsfig}

\newcommand{\AmS}{{\protect\the\textfont2
  A\kern-.1667em\lower.5ex\hbox{M}\kern-.125emS}}
\title{The shape function in field theory}
\author{U. Aglietti \address{CERN-TH, Geneva, Switzerland}}

\begin{document}

\begin{abstract}
The shape function describes (non-perturbative) Fermi motion effects in
semi-inclusive heavy flavour decay. Its renormalization properties are
substantially dependent on the kind of ultraviolet regulator used. For
example, the double logarithm that appears at one loop is larger
by a factor $2$ in dimensional regularization than in lattice
regularization. We show that factorization of long-distance effects inside
the shape function is achieved with any regulator considered.
\end{abstract}

\maketitle

\section{Introduction}

In general, the shape function \cite{shapefunction} is introduced to
describe the decay of a hadron $H_{Q}$ containing a heavy flavour $Q$ into an
inclusive hadronic state $X$ with a large energy and a small invariant mass,
plus non-QCD partons, i.e. 
\begin{equation}
H_{Q}\rightarrow X+{\rm non-QCD~\,partons~~ }\left( m_{X}\ll
E_{X}\right) .  \label{base}
\end{equation}
Specifically, the selected kinematics is 
\begin{equation}
\frac{m_{X}^{2}}{E_{X}}\sim O\left( \Lambda _{QCD}\right) ,\qquad E_{X}\gg
\Lambda _{QCD}.  \label{conditio}
\end{equation}
In a more formal language, the conditions (\ref{conditio}) correspond to the
limit 
\begin{equation}
E_{X}\rightarrow \infty ,\quad m_{X}\rightarrow \infty ,\quad {\rm %
with\,\,\,\,}\frac{m_{X}^{2}}{E_{X}}\rightarrow {\rm const.}  \label{limite}
\end{equation}
The limit (\ref{limite}) implies the infinite mass limit of the heavy
flavour, 
\begin{equation}
m_{Q}\rightarrow \infty ,
\end{equation}
as $m_{Q}\geq E_{X}.$ The heavy quark can then be treated in the Heavy Quark
Effective Theory  (HQET) \cite{hqet}. The divergence of $m_{X}$ - even
though 
it is slower than the one of $E_{X}$ - implies that the final
hadronic state can be replaced by a partonic one and that perturbation
theory (PT) can be applied; however, the decay has also a
non-perturbative component - to be factorized in the shape function - which
we introduce with the following model.

We identify a hard subprocess in (\ref{base}) consisting in the
fragmentation of the heavy flavour $Q$, 
\begin{equation}
Q\rightarrow \widehat{X}+{\rm non-QCD~\,partons.}  \label{partonico}
\end{equation}
$\widehat{X}$ differs from $X$ in that it does not contain the light
valence
quark(s) of $H_{Q}$. The momentum of $Q$ can be written as \cite{hqet}
\footnote{
We use $m_{H}$ instead of $m_{Q}$ to define the momentum (\ref{peso}); this
amounts to a shift in the range of $\ k_{+}^{\prime }$ \cite{conguido}.} 
\begin{equation}
p_{Q}=m_{H}v+k^{\prime },  \label{peso}
\end{equation}
where $m_{H}$ is the mass and 
\begin{equation}
v^{\mu }=\left( 1;0,0,0\right) 
\end{equation}
is the velocity of $H_{Q}$, 
which we take at rest without any loss of generality. 
The momentum of the light degrees of freedom in $H_{Q}$ is $-k^{\prime }$.
The point $k^{\prime }=0$ is the elastic one where all the
initial light partons have soft momenta. It is natural to assume that \cite
{hqet,memasse} 
\begin{equation}
k_{+}^{\prime }\sim k_{-}^{\prime }\sim k_{\perp }^{\prime }\sim O\left(
\Lambda _{QCD}\right) .  \label{basic}
\end{equation}
The distribution of the $k^{\prime }$-momenta is clearly non-perturbative
and it is related to the well-known Fermi motion of the heavy quark inside
the hadron \cite{ACCMM}. Let us denote by $q^{\mu }$ the momentum carried
away by the non-QCD partons, taken along the $+z$ axis; the final hadronic
system flies along the minus direction. $\widehat{X}$ \ has a mass 
\begin{equation}
m_{\widehat{X}}^{2}=\left( Q+k^{\prime }\right)
^{2}=m_{X}^{2}+2E_{X}\,k_{+}^{\prime }+\frac{m_{X}^{2}}{2E_{X}}k_{-}^{\prime
}+k^{\prime 2},  \label{massinv}
\end{equation}
where $Q^{\mu }$ is the momentum available to the final partons 
\footnote{%
The light-cone components are defined as $a_{\pm }\equiv a_{0}\pm a_{3}$.}, 
\begin{eqnarray}
Q &\equiv &m_{H}v-q=\left( m_{H}-q_{0};0,0,-q_{3}\right)   \nonumber \\
&=&\left( E_{X};0,0,-\sqrt{E_{X}^{2}-m_{X}^{2}}\right)   \nonumber \\
&\cong &E_{X}\left( v_{-}+\frac{m_{X}^{2}}{4E_{X}^{2}}v_{+}\right) ,
\end{eqnarray}
with $v_{\pm }=\left( 1;0,0,\pm 1\right) $. The sizes of the terms in the
third member of eq. (\ref{massinv}) are 
\begin{eqnarray}
2E_{X}\,k_{+}^{\prime } &\sim &O\left( E_{X}\Lambda _{QCD}\right) ,\, 
\nonumber \\
\,\frac{m_{X}^{2}}{2E_{X}}k_{-}^{\prime } &\sim &k^{\prime 2}\sim O\left(
\Lambda _{QCD}^{2}\right) .  \label{stime}
\end{eqnarray}
We now perform two different approximations:

\begin{enumerate}
\item  We linearize the problem by dropping the $k^{\prime 2}$ term in eq. (\ref
{massinv}), i.e. we describe the final hadronic system with the $HQET$
or, in geometrical language, with a Wilson line {\it off} the light cone;

\item  We drop the term $m_{X}^{2}/\left( 2E_{X}\right) k_{-}^{\prime }$ in
the last member of eq.(\ref{massinv}). $Q^{\mu }$ is replaced
with a vector lying exactly on the light cone, in the minus direction. 
Neglecting these mass (virtuality) effects, the final hadronic system is described by
the LEET \cite{firstleet} or, in geometrical language, by a Wilson line 
{\it on} the light cone.
\end{enumerate}
Therefore, we have 
\begin{equation}
m_{\widehat{X}}^{2}\simeq m_{X}^{2}+2E_{X}\,k_{+}^{\prime }.  \label{finale}
\end{equation}
Equation (\ref{finale}) is the main result of this section. Let us comment on it. 
$m_{\widehat{X}}^{2}$ depends on a single light-cone component, 
$k_{+}^{\prime }$
in our reference frame. Because of eqs. (\ref{conditio}) and (\ref{basic}),
the two terms on the r.h.s. of eq.(\ref{finale}) are of the same order:  
thus $m_{\widehat{X}}^{2}$ is affected in a substantial way by
the distribution of the $k_{+}^{\prime }$ momenta and cannot 
be considered constant.

The rate of the elementary process (\ref{partonico}) contains perturbative 
corrections of double-logarithmic kind: 
\begin{equation}
\alpha _{S}^{n}\left( \frac{\log ^{k}\left[ 1-x\right] }{1-x}\right)
_{+}\qquad \left( 0\leq k\leq 2n-1\right) ,  \label{suda}
\end{equation}
where \footnote{
The plus-distribution $P\left( x\right) _{+}$ is defined on test functions $%
f\left( x\right) $ in the unit interval $x\in \lbrack 0,1]$ as $%
\int_{0}^{1}P\left( x\right) \left[ f\left( x\right) -f\left( 1\right) %
\right] dx.$} 
\begin{equation}
x\equiv 1-\frac{m_{\widehat{X}}^{2}}{E_{X}^{2}}\qquad \qquad (E_{\widehat{X}%
}\cong E_{X}).
\end{equation}
Since $m_{X}^{2}$ and $m_{\widehat{X}}^{2}$ are of the same order of
magnitude, the limit (\ref{limite}) implies that the threshold region (also
called large-$x$ region) is approached: 
\begin{equation}
x\rightarrow 1^{-}.
\end{equation}
The perturbative corrections (\ref{suda}) --- enhanced in the threshold region
--- are large and radically modify the tree-level distribution \cite{cattren}.
The physical distribution for (\ref{base}) is obtained by convoluting the
perturbative corrections of the form (\ref{suda}) with the primordial $%
k_{+}^{\prime }$-distribution: this is the way non-perturbative effects
enter the game. We conclude that the process has a substantial
non-perturbative component related to the $\,k_{+}^{\prime }$-distribution.

Let us also present another way to establish the non-perturbative component
in the decay (\ref{base}), which is a critical analysis of dynamics using
only perturbation theory. We neglect all the binding effects (confinement,
Fermi motion, etc.) and consider an isolated on-shell heavy quark $Q$, i.e.
we take $k^{\prime }=0$ \footnote{%
The physical process and the subprocess coincide in this case.}. $Q$ decays
into a massless quark $q$ plus non-QCD partons: 
\begin{equation}
Q\rightarrow q+{\rm non-QCD~\,partons.}
\end{equation}
We now consider the emission of a soft gluon, with momentum components of
the order of the QCD scale: 
\begin{equation}
k_{+}\sim k_{-}\sim k_{\perp }\sim O\left( \Lambda _{QCD}\right) .
\label{soft}
\end{equation}
The invariant mass of the final hadronic state is 
\begin{equation}
m_{X}^{2}=(p_{q}+k)^{2}\simeq 2E_{X}\,k_{+}\sim O\left( E_{X}\,\Lambda
_{QCD}\right) .\,
\end{equation}
An invariant mass of the order of (\ref{conditio}), i.e. rather large, is
generated by the emission of a very soft gluon. A kinematical amplification
by a factor $E_{X}$ has occurred as the kinematics goes from time-like
to light-like with the fragmentation of the heavy flavour. Higher-order
perturbative corrections replace the bare coupling with the running coupling
evaluated at the transverse gluon momentum \cite{veneziano}, 
\begin{equation}
\alpha _{S}\rightarrow \alpha _{S}\left( k_{\perp }\right) .
\end{equation}
For a soft gluon with the momentum (\ref{soft}), the coupling is evaluated
close to the Landau pole, indicating the presence of non-perturbative
effects. We conclude again that the decay (\ref{base}) has the
non-perturbative component identified before.

Let us observe that the linearization introduced in $1.$ and the light-cone
limit introduced in $2.$ are not valid approximations for a hard collinear
gluon;
by this we mean a gluon with momentum components of order 
\[
k_{-}\sim O\left( E_{X}\right) ,\quad k_{+}\sim O\left( \Lambda
_{QCD}\right) {\rm \,} 
\]
and, in order to have $k^{2}\sim 0$: 
\begin{equation}
k_{\perp }\sim O\left( \sqrt{E_{X}\,\Lambda _{QCD}}\right).
\label{hardcoll}
\end{equation}
For such a gluon, all the terms in the last member of eq. (\ref{massinv})
are of the same order. Its contribution can be considered a short-distance
effect, since the transverse momentum (\ref{hardcoll}) is substantially
larger than that of a soft gluon (\ref{soft}) and the related coupling
constant is in the perturbative region.

According to the above discussion, it is natural to identify the
non-perturbative component in (\ref{base}) as the following matrix element: 
\begin{equation}
f\left( k_{+}\right) \equiv \langle H_{Q}\left( v\right) |h_{v}^{\dagger
}\delta \left( k_{+}-iD_{+}\right) h_{v}|H_{Q}\left( v\right) \rangle ,\,
\end{equation}
where $h_{v}$ is a field in the HQET with velocity $v$. This function
gives the probability that the heavy quark in the hadron has a plus
virtuality equal to $k_{+}$, independently from the other components. In the
following section we will critically analyse how the above matrix element
factorizes the non-perturbative effects. We already found that the hard
collinear region (\ref{hardcoll}) cannot be described by the shape function
but can consistently be incorporated into a coefficient function \footnote{%
The coefficient function is also called jet factor, collinear factor,
matching constant, hard factor and short-distance cross-section.} \cite
{cattren}.

\section{The shape function in various regularizations}

In order to avoid distributions and to deal only with ordinary functions, it
is convenient to consider the light-cone function \cite{megiulia} 
\begin{equation}
F\left( k_{+}\right) \equiv \langle H_{Q}|h_{v}^{\dagger }\frac{1}{%
k_{+}-iD_{+}+i\epsilon }h_{v}|H_{Q}\rangle ,
\end{equation}
from which the shape function is obtained by taking the imaginary part, 
\begin{equation}
f\left( k_{+}\right) =-\frac{1}{\pi }{\rm Im}~F\left( k_{+}\right) .
\end{equation}
We want to study if the light-cone function factorizes the non-perturbative
effects by performing a perturbative computation. This involves $3$ steps:

\begin{enumerate}
\item  To replace the hadronic light-cone function with a partonic
light-cone function \footnote{%
The same symbol $F\left( k_{+}\right) $ is used for the two different matrix
elements, as this should not cause confusion.}, 
\begin{equation}
F\left( k_{+}\right) =\langle Q|h_{v}^{\dagger }\frac{1}{k_{+}-iD_{+}+i%
\epsilon }h_{v}|Q\rangle ,
\end{equation}
and to perform a perturbative computation. We take the heavy quark $Q$ with
the momentum (\ref{peso}), i.e off-shell by $k^{\prime }$;

\item  To perform the same perturbative computation --- with the same external
states --- in the original high-energy theory, i.e. full QCD;

\item  To compare the results to see if the difference is a 
short-distance effect or not.
\end{enumerate}

All this procedure has a meaning if we accept the following assumption: the
long-distance effects of {\it perturbative} kind --- i.e. the leading infrared
logarithms --- are able to trace the long-distance effects of {\it %
non-perturbative} nature. Therefore, if two matrix elements have the {\it %
same} {\it perturbative} long-distance contributions, they manifest also the 
{\it same} {\it non-perturbative} long-distance effects if computed with a
non-perturbative technique, such as lattice QCD \footnote{%
An exception to this rule seems to be the observed difference of the
fragmentation functions of the heavy flavours, as extracted from electron and
proton collisions \cite{veroe}.}.

At the tree level, we have the expected result 
\begin{equation}
F(k_{+})^{tree}=\frac{1}{k_{+}^{\prime }-k_{+}+i\epsilon },
\end{equation}
i.e. 
\begin{equation}
~~~~f(k_{+})^{tree}=\delta \left( k_{+}^{\prime }-k_{+}\right) .
\end{equation}
The shape function is a spike when its argument matches the plus virtuality
of the external state, $k_{+}=k_{+}^{\prime }$, independently of the other
components, $k_{-}^{\prime }$ and $k_{\perp }^{\prime }$. We now consider
the double logarithm that appears at one-loop.

Let us first discuss the case of a simple regularization cutting the space
momenta: 
\begin{equation}
|\overrightarrow{l}|<\Lambda _{S},\qquad -\infty <l_{0}<+\infty .
\label{semprelei}
\end{equation}
Since infrared logarithms are associated to quasi-real configurations, for
which $l_{0}\sim |\overrightarrow{l}|$, we expect this regularization to
give the same double logarithm as lattice regularization (after continuation
from Euclidean to Minkowski space). On the lattice, all the momentum
components are cut off,
\begin{equation}
|l_{\mu }|<\frac{\pi }{a},
\end{equation}
where $a$ is the lattice spacing. The result of the
computation done with the
   regularization (\ref
{semprelei}) is    \cite{megiulia} 
\begin{equation}
F(k_{+})^{\Lambda _{S}}=\frac{1}{k+i0}\ \left( -\frac{a}{2}\right) \ \log
{}^{2}\left( \frac{\Lambda _{S}}{-k-i0}\right) ,  \label{nostra}
\end{equation}
where we defined the usual combination 
\begin{equation}
k\equiv k_{+}^{\prime }-k_{+}.
\end{equation}
and $a\equiv \alpha _{S}C_{F}/\pi .$

The result of the computation done with
Dimensional
Regularization 
  (DR)        is    \cite
{ks,megiulia,luke} \footnote{%
This           result is also in \cite{mannel},
but we disagree with the
renormalization procedure \cite{megiulia}.} 
\begin{eqnarray}
F_{B}(k_{+}) &=&\frac{1}{k+i0}\left( -\frac{a}{2}\right) \Gamma (1+\epsilon )
\nonumber \\
&&\frac{\Gamma (1+2\epsilon )\Gamma (1-2\epsilon )}{\epsilon ^{2}}\left( 
\frac{\mu }{-k-i0}\right) ^{2\epsilon }  \nonumber \\
&=&\frac{1}{k+i0}\ a\left[ -\frac{1}{2\epsilon ^{2}}-\frac{1}{\epsilon }\log
{}\left( \frac{\mu }{-k-i0}\right) \right. \   \nonumber \\
&&\left. -\log {}^{2}\left( \frac{\mu }{-k-i0}\right) \right] ,  \label{DRF}
\end{eqnarray}
where $\epsilon \equiv 2-D/2$ (with $D$ the space-time dimension)
and $\mu $ is
the regularization scale \footnote{%
Actually, in the last member of (\ref{DRF}), $\mu ^{2}$ should be replaced
by $\mu ^{2}4\pi $ exp$\left[ -\gamma _{E}\right] $ ($\gamma _{E}$ is the
Euler constant), even though this rescaling does not affect DLA results.}.
We find a double pole of UV nature
because infrared singularities
(soft and
collinear) are completely regulated by the light-quark leg being
off-shell. The first problem we encounter       is how to renormalize the
above expression. 
The point is that, to obtain a finite result, one has to subtract not only
the double pole, whose coefficient is just a number, but also the simple
pole, which has a $\log \mu /k$ coefficient \cite{taylor,luke}.
It seems that the subtraction of the
simple pole cannot be performed with a local counter-term, as standard
textbooks on renormalization teach one should do. However, if we blindly
subtract all the poles, we obtain 
\begin{equation}
F(k_{+})^{"ren"}=\frac{1}{k+i0}\ \left( -a\right) \ \log {}^{2}\left( \frac{%
\mu }{-k-i0}\right) .  \label{DRFR}
\end{equation}
We see that the coefficient of the double logarithm is $2$ times larger than
in eq.
(\ref{nostra}).

Even though we are not able to justify the renormalization procedure, we
believe that the above result is correct. A hint in favour of this
conjecture is obtained by
repeating the same computation in the bare theory at
finite ultraviolet         (UV)        cut-off, using a regularization
similar to DR, i.e. cutting only the transverse momenta and not the
longitudinal ones \footnote{%
The loop measure in DR is indeed regulated as $\int d^{D}l=1/2\int
dl_{+}dl_{-}d^{D-2}l_{\perp }$.}: 
\begin{equation}
|\overrightarrow{l}_{\perp }|<\Lambda _{\perp },\qquad -\infty
<l_{+},\,l_{-}<\infty .
\end{equation}
The unbounded integration over the longitudinal components does not give
rise to UV          divergences  and we obtain the result \cite{megiulia} 
\begin{equation}
F(k_{+})^{\Lambda _{\perp }}=\frac{1}{k+i0}\ \left( -a\right) \ \log
{}^{2}\left( \frac{\Lambda _{\perp }}{-k-i0}\right) .  \label{LambdaT}
\end{equation}
The transposition of symbols should be clear: the renormalization point $\mu 
$ is replaced by the  UV cut-off $\Lambda _{\perp }$. As anticipated, there 
is   agreement between the two results (\ref{DRFR}) and (\ref{LambdaT}).

The origin of the factor $2$ in the difference of the double logarithms is
easily seen by
considering the integrals 
\begin{eqnarray*}
I_{\Lambda _{S},\Lambda _{\perp }} &\equiv &\int_{0}^{\infty }\frac{%
d\epsilon }{\epsilon }\int_{0}^{\epsilon ^{2}}\frac{dl_{\perp }^{2}}{%
l_{\perp }^{2}}\theta \left( l_{\perp }^{2}-\epsilon k_{+}\right)  \\
&&\quad \quad \,\,\theta \left( \Lambda _{S}-\epsilon \right) ,\,\theta
\left( \Lambda _{\perp }^{2}-l_{\perp }^{2}\right) .
\end{eqnarray*}
A straightforward
computation indeed gives:
\[
I_{\Lambda _{S}}=\frac{1}{2}\log ^{2}\frac{\Lambda _{S}}{k_{+}},\quad
I_{\Lambda _{\perp }}=\log ^{2}\frac{\Lambda _{\perp }}{k_{+}}.
\]
Let us analyse the domains of the $2$ integrals in the plane $\left(
\epsilon ,l_{\perp }\right) $. For $I_{\Lambda _{S}}$, the integration
region is a triangle-like domain $D_{S}$ limited by the curves 
\[
l_{\perp }=\sqrt{k_{+}\epsilon },\quad l_{\perp }=\epsilon \quad {\rm and}%
\quad \,\epsilon =\Lambda _{S}\qquad \left( D_{S}\right) .
\]
The energies and the transverse
momenta in $D_{S}$ are in the ranges (see fig.1) 
\[
k_{+}<\epsilon <\Lambda _{S},\quad \quad k_{+}<l_{\perp }<\Lambda
_{S}~~~~~~~~~~~~(D_S).
\]
Transverse momenta do become as small as $k_{+}\sim O\left( \Lambda
_{QCD}\right) $, entering the non-perturbative region.
For $I_{\Lambda _{\perp }}$%
, the integration region $D_{\perp }$ has again a triangle-like   shape and
it is limited by the curves 
\[
l_{\perp }=\sqrt{k_{+}\epsilon },\quad l_{\perp }=\epsilon \quad {\rm %
and\quad }l_{\perp }=\Lambda _{\perp }\qquad \left( D_{\perp }\right) .
\]
Assuming $\Lambda _{\perp }=\Lambda _{S},$ we see  that $D_{\perp }$
contains $D_{S}$ plus another region $\Delta D$ (see fig. 2),
\[
D_{\perp }=D_{S}\cup \Delta D.
\]
In the  ``new''  region $\Delta D$ 
\[
\sqrt{k_{+}\Lambda _{\perp }}<l_{\perp }<\Lambda _{\perp },\qquad \Lambda
_{\perp }<\epsilon <\frac{\Lambda _{\perp }^{2}}{k_{+}}\qquad \left( \Delta
D\right). 
\]
The integration over $\Delta D$ gives a double logarithmic
contribution equal to the one coming from the integration over $D_{S}$: that
is the origin of the factor $2$ in the difference between $I_{\Lambda _{S}}$
and $I_{\Lambda _{\perp }}$. The contribution coming from $\Delta D$ is a
short-distance contribution because it always involves rather large
transverse momenta, 
\begin{equation}
 l_{\perp }\geq \sqrt{k_{+}\Lambda _{\perp
}}\gg
\Lambda _{QCD}~~~~~~~~~(\Lambda _{\perp }\gg \Lambda _{QCD} ).
\end{equation}
 Let us observe also
that in $\Delta D$ the energy $\epsilon $ of the gluon can become much
larger than the  UV  cutoff, $\epsilon \gg \Lambda _{\perp }$. One usually
relates the cutoff to the hard scale $E_{X}$ in the full QCD process,
which, on the contrary, sets the upper bound for the gluon energies: $%
\epsilon \leq E_{X}$. It is clear that we are dealing with highly unphysical
regularization scheme effects.

We now compare with the QCD rate, from which a full QCD light-cone
function can be defined \cite{megiulia}, given by 
\begin{equation}
F(k_{+})^{QCD}=\frac{1}{k+i0}\ \left( -\frac{a}{2}\right) \ \log
{}^{2}\left( \frac{E_{X}}{-k-i0}\right) .  \label{full}
\end{equation}
It is clear that $\Lambda _{\perp }$ or $\mu $ is replaced by $E_{X}$. It is
immediate to see that the double logarithms match for the $\Lambda _{S}$%
-regularization, while they do not match for the $\Lambda _{\perp }$%
-regularization. In the latter case the ``spurious'' contribution from $%
\Delta D$ can be subtracted by means of the coefficient function, which
includes also   ``true'' hard collinear effects, finite terms, etc. While in
the $\Lambda _{S}$ case the coefficient function contains at most a single
logarithm of $k_{+}$, in the $\Lambda _{\perp }$ case it contains also a
double logarithm of $k_{+}.$

\section{Conclusions}

We tried to describe in a way as transparent as possible
the problem of factorization of non-perturbative
effects in the decay (\ref{base}) by means of the shape function. 
The starting point of our analysis is the
observation that the shape function exhibits a different double 
logarithm at one loop, depending on the kind of UV regularization adopted 
\cite{megiulia}. In a regularization such as
DR --- cutting only transverse
loop momenta ---
the double logarithm has a coefficient $2$ times larger than
the one derived with a lattice-like regularization. We have explicitly shown
that the difference between
the double logarithms is related to the integration
over a region of large transverse momenta, so it is a short-distance
effect.
The QCD rate has a double logarithm equal to the one derived in
lattice-like regularization. This implies that the coefficient function
in lattice-like
regularization contains at most a single logarithm of $k_{+}$ (the infrared
scale of the problem),
while in DR it contains also a double logarithm of $%
k_{+}$. In both regularizations, however, non-perturbative effects are
factorized inside the shape function.

In general, we observe that there are UV regularization effects similar to
those found for the shape function in other   operators incorporating Wilson
lines on the light cone \cite{futuro}.

\begin{center}
\bigskip Acknowledgements
\end{center}

I would like to thank S. Catani, A. Hoang and G. Korchemsky for discussions.

\newpage 

\noindent
Figure captions

$~~$

\noindent
Fig.1 (above): Integration region for the soft gluon in the plane
$(\epsilon,l_{\perp})$ with the $\Lambda_S$ regularization.
We have taken $k_+=0.3$
GeV and $\Lambda_S=1$ GeV.

$~~$

\noindent
Fig.2 (below): Integration region for the soft gluon in the plane
$(\epsilon,l_{\perp})$ with the $\Lambda_{\perp}$ regularization. We have
taken $k_+=0.3$ GeV and $\Lambda_{\perp}=1$ GeV. The domain $D_S$ is to
the left of the dashed line, $\Delta D$ is to the right.


\begin{figure}[htb]
\centerline{\mbox{\epsfxsize=10cm\epsffile{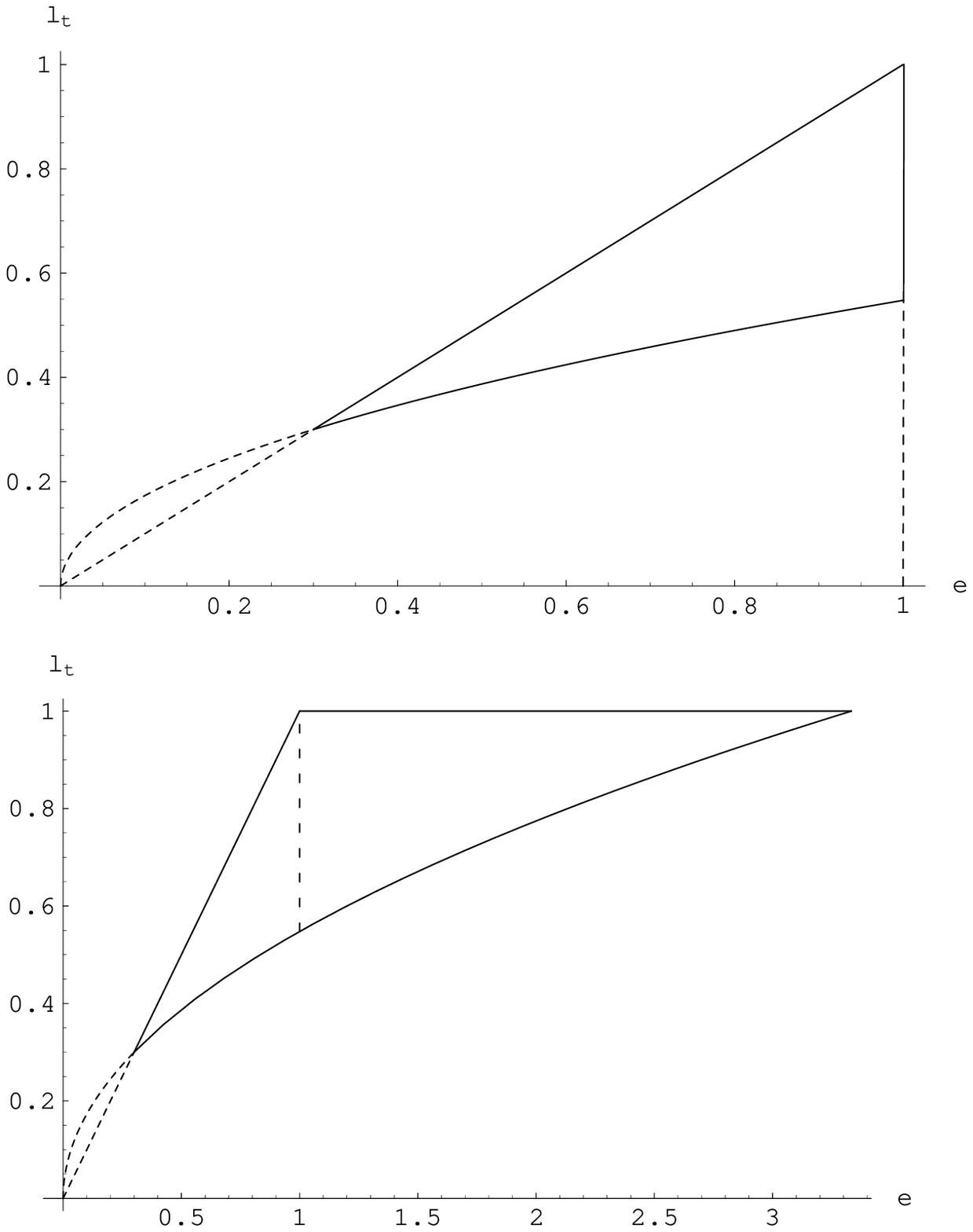}}}
\end{figure}


\begin{thebibliography}{99}
\bibitem{shapefunction}  I. Bigi, M. Shifman, N. Uraltsev and A. Vainshtein,
Phys. Rev. Lett. 71:496 (1993); A. Manohar and M. Wise, Phys. Rev. D 49:1310
(1994); M. Neubert, Phys. Rev. D 49:3392 (1994), ibid. D 49:4623 (1994).

\bibitem{hqet}  F. Bloch and A. Nordsieck, Phys. Rev. 52:54 (1937); L. Foldy
and S.~Wouthuysen, Phys. Rev. 78:29 (1950); D. Yennie, S. Frautschi and
H.~Suura, Ann. Phys. 13:379 (1961); E. Eichten and F. Feinberg, Phys. Rev.
D~23:2724 (1981); N. Isgur and M. Wise, Phys. Lett. B 237:527 (1990); H.
Georgi, Phys. Lett. B 240:447 (1990); F. Hussain et al., Phys. Lett. B
249:295 (1990); U. Aglietti and S. Capitani, Nucl. Phys. B 432:315-336\
(1994).

\bibitem{conguido}  U. Aglietti et al., Phys. Lett. B 432:411-420 (1998).

\bibitem{memasse}  U. Aglietti, Phys. Lett. B 281:341 (1992).

\bibitem{ACCMM}  G. Altarelli et al., Nucl. Phys. B 208:365 (1982).

\bibitem{firstleet}  M. Dugan and B. Grinstein, Phys. Lett. B 255:583
(1991); U. Aglietti, Phys. Lett. B 292:424 (1992).

\bibitem{cattren}  J. Kodaira and L. Trentadue, SLAC preprint SLAC-PUB-2934
(1982) and Phys. Lett. B 112:66 (1982), B 123:335 (1983); S. Catani and L.
Trentadue, Nucl. Phys. B 327:323 (1989).

\bibitem{veneziano}  D. Amati et al.                                                     Nucl. Phys. B 173:429 (1980).

\bibitem{megiulia}  U. Aglietti and G. Ricciardi, Phys. Lett. B 466:313
(1999); CERN-TH/2000-071 and DSF-T/2000-7, hep-ph/0003146, to appear in
Nucl. Phys. B.

\bibitem{veroe}  G. Martinelli, private communication; S. Frixione and M.
Mangano in \ ``Proc.              Workshop on Standard Model Physics (and
More) at the LHC'',        Resport CERN 2000-004, p. 254.

\bibitem{ks}  G.~Korchemsky and G. Sterman, Phys. Lett. B 340:96 (1994); A.
Grozin and G. Korchemsky, Phys. Rev. D 53, 1378 (1996); R. Akhoury and
I.~Rothstein, Phys. Rev. D 54:2349 (1996); A. Leibovich and I.~Rothstein,
Phys. Rev. D 61:074006 (2000); A. Leibovich, I. Low and I.~Rothstein, Phys.
Rev. D 61:053006 (2000); 

\bibitem{luke}  C. Bauer, S. Fleming and M. Luke, UTPT-00-03, hep/ph0005275.

\bibitem{mannel}  C. Balzereit, W. Kilian and T. Mannel, Phys. Rev. D
58:114029 (1998).

\bibitem{taylor}  A. Andrasi and J. C. Taylor, Nucl. Phys. B 350:73-81
(1991).

\bibitem{futuro} U. Aglietti, work in progress.
\end{thebibliography}
\end{document}